\documentclass{llncs}

\usepackage{latexsym}
\usepackage{graphicx}
\usepackage{epsfig}
\usepackage{amsmath,amssymb,amsfonts}
\usepackage{enumitem}

\usepackage{color}
\usepackage{textcomp}
\usepackage{tabularx}

\graphicspath{{figures/}} 

\newcommand{\remove}[1]{}

\long\def\invis#1{}

\begin{document}
\title{Re-embedding a 1-Plane Graph  into a  Straight-line Drawing
in Linear Time
 \thanks {Research supported by ARC Future Fellowship and ARC Discovery Project
 DP160104148. This is an extended abstract. For a full version with omitted proofs, see~\cite{TR}.}
}

\author{Seok-Hee Hong}

\institute{
School of Information Technologies \\
University of Sydney, Australia \\
\email{shhong@it.usyd.edu.au}
}

\author{Seok-Hee Hong\inst{1}
\and Hiroshi Nagamochi\inst{2}}

\institute{
University of Sydney, Australia \\
\email{seokhee.hong@sydney.edu.au}
\ \\
\and
Kyoto University, Japan\\
\email{nag@amp.i.kyoto-u.ac.jp}
}

\date{}
\maketitle

\begin{abstract}
Thomassen characterized
some 1-plane embedding as the forbidden configuration such that
a given 1-plane embedding of a graph is drawable in straight-lines
  if and only if it does not contain
the  configuration
[C. Thomassen, Rectilinear drawings of graphs,
J. Graph Theory, 10(3),  335-341,  1988].

In this paper, we characterize
some 1-plane embedding as the forbidden configuration such that
a given 1-plane embedding of a graph can be re-embedded into
 a   straight-line drawable 1-plane embedding of the same graph
  if and only if it does not contain the  configuration.
Re-embedding of a 1-plane embedding preserves the same set of pairs of
crossing edges.
We give a linear-time algorithm for finding a
straight-line drawable 1-plane re-embedding or
the forbidden configuration.
\end{abstract}

\section{Introduction}\label{se:introduction}

Since the 1930s, a number of researchers have investigated {\em planar} graphs.
In particular, a beautiful and classical result, known as \emph{F\'{a}ry's Theorem}, asserts that
every plane graph admits a {\em straight-line drawing}~\cite{Fary}.
Indeed, a straight-line drawing is the most popular drawing convention in Graph Drawing.

More recently, researchers have investigated \emph{1-planar graphs}
 (i.e., graphs that can be embedded in the plane with at most one crossing per edge), introduced by Ringel~\cite{Ringel}.
Subsequently, the structure of 1-planar graphs has been investigated~\cite{FM,PT}.
In particular, Pach and Toth~\cite{PT} proved that a 1-planar graph
with $n$ vertices has at most $4n-8$ edges, which is a tight upper bound.
Unfortunately, testing the 1-planarity of a graph is NP-complete~\cite{Bodlaender,Mohar},
however linear-time algorithms are available for special subclasses of 1-planar graphs~\cite{Franz,Pascal,GD13}.

\begin{figure}[htbp]
 \centering
 \includegraphics[width=.8\columnwidth]{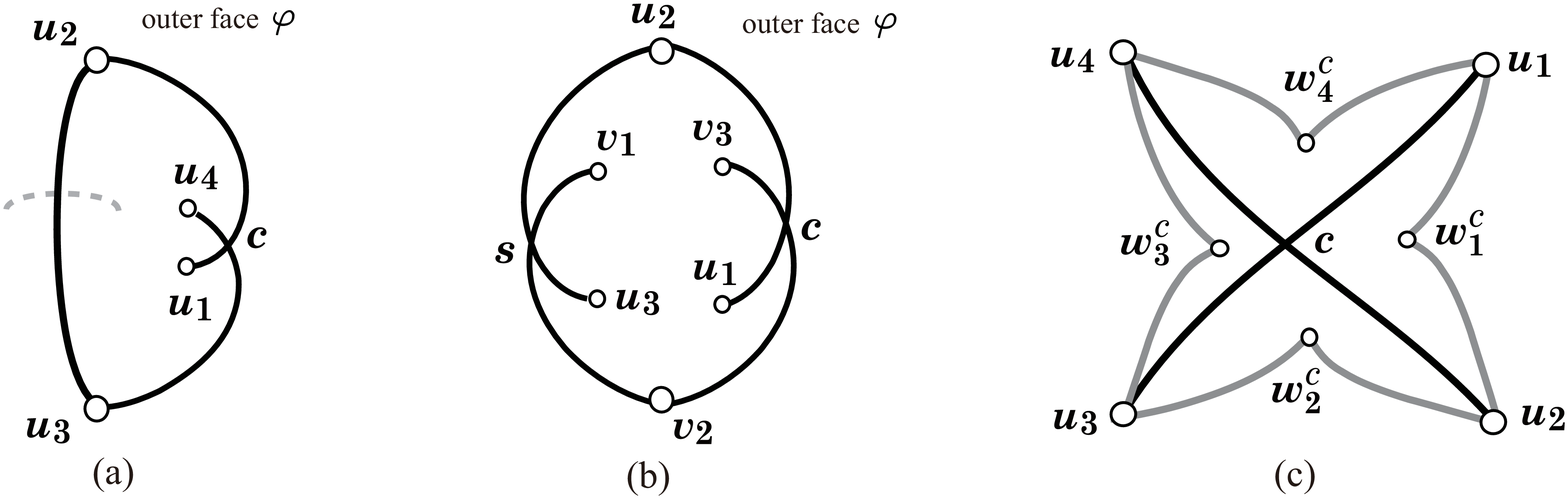}
    \caption{(a)  B-configuration with three edges
     $u_1u_2,u_2u_3$ and $u_3u_4$ and one crossing $c$
    made by an edge pair $\{u_1u_2, u_3u_4\}$, where edge $u_2u_3$ may have a crossing when
       the configuration is part of a 1-plane embedding;
     (b)   W-configuration with four edges
           $u_1u_2$, $u_2u_3$, $v_1v_2$ and $v_2v_3$ and two crossings $c$ and $s$
    made by   edge pairs $\{u_1u_2, v_2v_3\}$ and $\{u_2u_3, v_1v_2\}$, where possibly
    $u_1=v_1$ and $u_3=v_3$;
    (c) Augmenting a crossing $c\in\chi$ made by edges $u_1u_3$ and $u_2u_4$ with a new
         cycle $Q_c=(u_1,w_1^c,u_2,w_2^c, u_3,w_3^c, u_4,w_4^c)$ depicted by gray lines. }
  \label{fi:BW-configuration}
\end{figure}

Thomassen~\cite{Thomassen} proved that every  1-plane graph
  (i.e.,  a 1-planar graph embedded with a given {\em 1-plane embedding}) admits a straight-line drawing
if and only if it does not contain any of two special 1-plane graphs,
called the {\em B-configuration} or {\em W-configuration}, see    Fig.~\ref{fi:BW-configuration}.

Recently, Hong et al.~\cite{HELP} gave an alternative constructive proof,
with a linear-time testing algorithm and a drawing algorithm.
They also showed that some 1-planar graphs need an exponential area with straight-line drawing.

We call a 1-plane embedding {\em straight-line drawable} ({\em SLD} for short)
 if it admits a  straight-line drawing,
i.e., it does not contain a B- or W-configuration by Thomassen~\cite{Thomassen}.
In this paper, we investigate a problem of ``re-embedding'' a given  non-SLD 1-plane embedding $\gamma$
 into an SLD 1-plane embedding $\gamma'$.
For a given 1-plane embedding $\gamma$ of a graph $G$,
we call another 1-plane embedding $\gamma'$ of $G$
 a {\em cross-preserving embedding} of $\gamma$ if
 exactly the same set of edge pairs make the same crossings in $\gamma'$.

More specifically,
we first characterize the {\em forbidden configuration} of 1-plane embeddings that
cannot admit an SLD cross-preserving 1-plane embedding.
Based on the characterization, we present a linear-time algorithm that
 either detects the   forbidden configuration in $\gamma$
or computes  an SLD cross-preserving 1-plane embedding $\gamma'$.

Formally, the main problem considered in this paper is defined as follows.

\medskip
 \noindent
 {\bf Re-embedding a 1-Plane Graph into a Straight-line Drawing }\\
 {\bf Input:}  A  1-planar graph $G$ and a 1-plane embedding $\gamma$ of $G$.\\
{\bf Output:}  Test whether $\gamma$ admits
  an SLD cross-preserving 1-plane embedding $\gamma'$,
and construct such an embedding $\gamma'$ if one exists, or report the forbidden configuration.\\

To design a  linear-time implementation of our algorithm in this paper,
we introduce a {\em rooted-forest representation of non-intersecting cycles}
and an efficient procedure of flipping subgraphs in a plane graph.
Since these data structure and procedure can be easily implemented,
it has advantage over the complicated decomposition of biconnected graphs into
triconnected components~\cite{Hopcroft3} or the SPQR tree~\cite{BT}.

\invis{
The paper is organized as follows.
Section~\ref{sec:plane_graph} makes a technical preparation to attain
the linear time complexity of our algorithm
by   introducing a forest representation of non-intersecting cycles
and an efficient implementation of a flip operation in a plane graph.
Section~\ref{sec:1-plane_graph} investigates
the structure of cycles  that can induce a B- or W-configuration
and defines a forbidden configuration to our problem.
Section~\ref{sec:biconnected} and Section~\ref{sec:connectivity-one} treat  the case
where the vertex-connectivity of the planarization of a 1-plane embedding $\gamma$
 is at least 2 and 1, respectively,
 where a linear-time algorithm is designed
  to detect  the   forbidden configuration in $\gamma$
or to construct   an SLD cross-preserving 1-plane embedding of $\gamma$.
Finally Section~\ref{sec:conclusion} makes some concluding remarks.
}

\section{Plane Embeddings and Inclusion Forests}\label{sec:plane_graph}

Let $U$ be a set of $n$ elements, and let $\mathcal{S}$ be
 a family of subsets $S\subseteq U$.
 We say that two subsets $S,S'\subseteq U$ are {\em intersecting}
 if none of $S\cap S'$, $S-S'$ and $S'-S$ is empty.
 We call $\mathcal{S}$ a  {\em laminar} if no two subsets in $\mathcal{S}$ are intersecting.
For a laminar $\mathcal{S}$,
the {\em inclusion-forest} of $\mathcal{S}$ is defined to be
a forest $\mathcal{I}=(\mathcal{S},{\cal E})$ of a disjoint union of rooted trees
  such that (i) the sets in $\mathcal{S}$ are regarded
as the vertices of $\mathcal{I}$, and
(ii)  a set $S$ is an ancestor of a set $S'$ in $\mathcal{I}$
if and only if
 $S'\subseteq S$.

\begin{lemma} \label{le:cyclic}
 For  a cyclic sequence $(u_1,u_2,\ldots,u_{\delta})$ of $\delta\geq 2$ elements,
 define an interval $(i,j)$ to be the set of elements $u_k$ with $i\leq k\leq j$ if $i\leq j$
 and $(i,j)=(i,\delta)\cup (1,j)$ if $i>j$.
 Let $\mathcal{S}$ be a set  of intervals.
A pair of two intersecting intervals in  $\mathcal{S}$ $($when $\mathcal{S}$
is not a laminar$)$ or  the inclusion-forest
   of  $\mathcal{S}$ $($when $\mathcal{S}$ is a laminar$)$
   can be obtained  in $O(\delta+|\mathcal{S}|)$ time.
\end{lemma}
%

Throughout the paper, a graph $G=(V,E)$ stands for a simple undirected graph.
The set of vertices and the set of edges  of a graph $G$ are denoted by
$V(G)$ and $E(G)$, respectively.
For a vertex $v$,
let $E(v)$   be the set of edges incident to  $v$,
  $N(v)$ be the set of neighbors of $v$,
  and $\mathrm{deg}(v)$ denote the degree $|N(v)|$ of $v$.
A simple path with end vertices $u$ and $v$ is called  a {\em $u,v$-path}.
For a subset  $X\subseteq V$, let $G-X$ denote
the graph obtained  from $G$ by removing   the vertices in $X$ together with the
edges in $\cup_{v\in X}E(v)$.

A {\em drawing} $D$ of a graph $G$ is a geometric representation of the
graph in the plane, such that each vertex of $G$ is mapped to a point in the plane, and each
edge of $G$ is drawn as a curve.
A  drawing $D$ of a graph $G=(V,E)$ is called {\em planar}
if there is no edge crossing.
A planar drawing $D$ of a graph $G$ divides the plane into several
connected regions, called {\em faces}, where a face enclosed by a closed walk of the graph is
called an {\em inner face} and the face  not enclosed by any closed walk
is called the {\em outer face}.

A planar drawing $D$ {\em induces} a plane embedding $\gamma$ of $G$, which is
defined to be a pair $(\rho, {\varphi})$ of the {\em  rotation system}
(i.e., the circular ordering of edges for each vertex)  $\rho$,
and the outer face ${\varphi}$ whose facial cycle $C_\varphi$ gives
the outer boundary of $D$.
Let $\gamma=(\rho, {\varphi})$ be a plane embedding of a   graph $G = (V, E)$.
We denote by $F(\gamma)$  the set of  faces in $\gamma$,
and by  $C_f$   the facial cycle determined by a face $f\in F$,
where we call a subpath of $C_f$   a {\em boundary path} of $f$.
For a simple cycle $C$ of $G$,
the plane is divided by $C$ in two regions,
 one containing only inner faces and the other containing  the outer area,
    where
     we say that  the former is {\em enclosed} by $C$  or the {\em interior} of $C$,
     while the latter is called the {\em exterior} of $C$.
We denote by $F_{\mathrm{in}}(C)$   the set of inner faces
   in the interior of  $C$,    by  $E_{\mathrm{in}}(C)$ the set of edges in $E(C_f)$ with $f\in F_{\mathrm{in}}(C)$,
   and by  $V_{\mathrm{in}}(C)$
 the set of end-vertices of edges in $E_{\mathrm{in}}(C)$.
 Analogously define $F_{\mathrm{ex}}(C)$, $E_{\mathrm{ex}}(C)$
 and $V_{\mathrm{ex}}(C)$  in the exterior of $C$.
 Note that  $E(C)=E_{\mathrm{in}}(C)\cap E_{\mathrm{ex}}(C)$
 and $V(C)=V_{\mathrm{in}}(C)\cap V_{\mathrm{ex}}(C)$.

For a subgraph  $H$   of $G$,
we define the embedding $\gamma|_H$ of $\gamma$ induced by
$H$ to be  a  sub-embedding of $\gamma$ obtained by removing
the vertices/edges not in $H$, keeping the same rotation system
 around each of the remaining vertices/crossings
and the same outer face.

\subsection{Inclusion Forests of Inclusive Set of Cycles}\label{sec:inclusion-forest}

In this and next subsections,
 let $(G,\gamma)$ stand for a plane embedding of  $\gamma=(\rho,\varphi)$
 of a biconnected simple graph $G=(V,E)$ with $n=|V|\geq 3$.

 Let $C$ be  a simple cycle in $G$.
We define the {\em direction} of $C$ to be
an ordered pair $(u,v)$ with $uv\in E(C)$ such that
the inner faces in $F_{\mathrm{in}}(C)$ appear on the right hand side
when we traverse $C$ in the order that we start $u$ and next visit $v$.
For simplicity,  we say that two simple cycles $C$ and $C'$ are {\em intersecting}
 if $F_{\mathrm{in}}(C)$ and $F_{\mathrm{in}}(C')$ are intersecting.

Let $\mathcal{C}$ be a set of simple cycles  in $G$.
We call $\mathcal{C}$ {\em inclusive} if no two cycles in $\mathcal{C}$
are intersecting, i.e., $\{F_{\mathrm{in}}(C) \mid C\in \mathcal{C}\}$ is a laminar.
When $\mathcal{C}$ is inclusive,
the {\em inclusion-forest} of $\mathcal{C}$ is defined to be
a forest $\mathcal{I}=(\mathcal{C},{\cal E})$ of a disjoint union of rooted trees
  such that: \\
(i) the cycles in $\mathcal{C}$ are regarded
as the vertices of $\mathcal{I}$, and \\
(ii)  a cycle $C$ is an ancestor of a cycle $C'$ in $\mathcal{I}$
if and only if  $F_{\mathrm{in}}(C')\subseteq F_{\mathrm{in}}(C)$.

Let $\mathcal{I}(\mathcal{C})$ denote the inclusion-forest  of $\mathcal{C}$.
For a vertex subset $X\subseteq V$, let $\mathcal{C}(X)$ denote the set of cycles $C\in \mathcal{C}$
such that $x\in V(C)$ for some vertex $x\in X$,
where we denote $\mathcal{C}(\{v\})$ by $\mathcal{C}(v)$ for short.

\begin{lemma} \label{le:cycle-inclusion}
For $(G,\gamma)$, let  $\mathcal{C}$ be a set of simple cycles of $G$.
Then  any of the following tasks can be executed
in $O(n+\sum_{C\in \mathcal{C}}|E(C)|)$ time. \vspace{-1mm}
\begin{enumerate}
\item[{\rm (i)}]
Decision of the directions of all cycles in  $\mathcal{C}$;
\item[{\rm (ii)}]
Detection of a pair of two intersecting cycles in  $\mathcal{C}$ when $\mathcal{C}$
is not inclusive,
and construction of  the inclusion-forests
   $\mathcal{I}(\mathcal{C}(v))$ for all vertices $v\in V$
 when $\mathcal{C}$ is  inclusive; and
\item[{\rm (iii)}]
Construction of  the inclusion-forest    $\mathcal{I}(\mathcal{C})$
 when $\mathcal{C}$ is  inclusive.
\end{enumerate}
\end{lemma}

\subsection{Flipping Spindles}\label{sec:flipping-process}

A simple cycle $C$ of $G$ is called a {\em spindle}
(or a {\em $u,v$-spindle}) of $\gamma$
if there are two vertices $u,v\in V(C)$ such that
no vertex in $V(C)-\{u,v\}$ is adjacent to any vertex
in the exterior of $C$, where we   call vertices $u$ and $v$
 the {\em junctions} of $C$.
Note that each of the two subpaths
of $C$ between $u$ and $v$ is a boundary path of some face in $F(\gamma)$.

Given $(G,\gamma)$,
we denote the rotation system around a vertex $v\in V$   by
  $\rho_{\gamma}(v)$.
 For a spindle $C$ in $\gamma$,
 let $J(C)$ denote the set of the two junctions of $C$.
 
{\em Flipping a  $u,v$-spindle $C$} means to modify
the rotation system of vertices in $V_{\mathrm{in}}(C)$
 as follows:\\
(i) For each vertex  $w\in V_{\mathrm{in}}(C)-J(C)$,
reverse the cyclic order of $\rho_{\gamma}(w)$; and \\
(ii) For each vertex  $u\in J(C)$, reverse
the order of subsequence of $\rho_{\gamma}(u)$   that consists
of vertices $N(u)\cap V_{\mathrm{in}}(C)$.

Every two distinct spindles $C$ and $C'$ in $\gamma$ are
 non-intersecting, and they always satisfy one of
 $E_{\mathrm{in}}(C)\cap E_{\mathrm{in}}(C')=\emptyset$,
 $E_{\mathrm{in}}(C)\subseteq E_{\mathrm{in}}(C')$,
 and $E_{\mathrm{in}}(C')\subseteq E_{\mathrm{in}}(C)$.
Let $\mathcal{C}$ be a set  of spindles in $\gamma$, which is always inclusive,
 and
let $\mathcal{I}(\mathcal{C})$ denote the inclusion-forest  of $\mathcal{C}$.

When we modify
the current embedding $\gamma$ by flipping each spindle in $\mathcal{C}$,
the resulting embedding $\gamma_{\mathcal{C}}$ is the same,
independent from the ordering of the flipping operation to the spindles,
since  for  two spindles  $C$ and $C'$ which share a common junction
 vertex $u\in J(C)\cap J(C')$,
 the sets $N(u)\cap V_{\mathrm{in}}(C)$ and $N(u)\cap V_{\mathrm{in}}(C')$
 do not intersect, i.e., they are disjoint or
 one is contained in the other.

Define the {\em depth}  of a vertex $v\in V$ in $\mathcal{I}$  to be
the number of spindles $C\in  \mathcal{C}$ such that $v \in V_{\mathrm{in}}(C)-J(C)$,
and denote by $\mathrm{p}(v)$ the parity of depth of vertex $v$,
i.e.,  $\mathrm{p}(v)=1$ if the depth is odd and
$\mathrm{p}(v)=-1$ otherwise.

For a vertex $v\in V$,
let $\mathcal{C}[v]$ denote the set of spindles $C\in \mathcal{C}$ such that
$v\in J(C)$, and
let $\gamma_{\mathcal{C}[v]}$ be the embedding obtained from $\gamma$ by
flipping all spindles in $\mathcal{C}[v]$.
Let $\mathrm{rev}\langle \sigma\rangle$ mean the reverse of a sequence $\sigma$.
Then we see that
$\rho_{\gamma_{\mathcal{C}}}(v)= \rho_{\gamma_{\mathcal{C}[v]}}(v)$
if  $\mathrm{p}(v)=1$; and
$\rho_{\gamma_{\mathcal{C}}}(v) =\mathrm{rev}\langle \rho_{\gamma_{\mathcal{C}[v]}}(v) \rangle$
  otherwise.
To obtain the embedding $\gamma_{\mathcal{C}}$ from the current embedding $\gamma$
by flipping each spindle in $\mathcal{C}$,  it suffices to show how to compute each of $\mathrm{p}(v)$
and $\rho_{\gamma_{\mathcal{C}[v]}}(v)$ for all vertices $v\in V$.

\begin{lemma} \label{le:spindle}
Given $(G,\gamma)$,
let $\mathcal{C}$ be a set  of spindles of $\gamma$.
Then any of the following tasks can be executed
in $O(n+\sum_{C\in\mathcal{C}}|E(C)|)$ time. \vspace{-1mm}
\begin{enumerate}
\item[{\rm (i)}]
Decision of parity $\mathrm{p}(v)$ of all vertices $v\in V$;  and
\item[{\rm (ii)}]
Computation of $\rho_{\gamma_{\mathcal{C}[v]}}(v)$
for  all vertices $v\in V$.
\end{enumerate}
\end{lemma}

\section{Re-embedding 1-plane Graph and Forbidden Configuration} \label{sec:1-plane_graph}

A  drawing $D$ of a graph $G=(V,E)$ is called a {\em 1-planar drawing}
if each edge has at most one crossing.
A 1-planar drawing  $D$ of   graph $G$ induces
 a {\em 1-plane embedding} $\gamma$ of $G$, which is
defined to be a tuple $(\chi,\rho,{\varphi})$ of  the {\em crossing system} $\chi$ of $E$,
 the rotation system $\rho$ of $V$,   and the outer face ${\varphi}$  of $D$.
The  {\em planarization} ${\cal G}(G,\gamma)$
of a 1-plane embedding  $\gamma$  of  graph $G$
is the plane embedding   obtained from $\gamma$
by regarding crossings also as   graph vertices, called crossing-vertices.
The set of vertices in ${\cal G}(G,\gamma)$  is given by $V\cup\chi$.
For a notational convenience,
we refer to a subgraph/face of  ${\cal G}(G,\gamma)$ as a  subgraph/face  in $\gamma$.

Let  $\gamma=(\chi,\rho,{\varphi})$ be a 1-plane embedding  of graph $G$.
We call another  1-plane embedding  $\gamma'=(\chi',\rho',{\varphi}')$  of graph $G$
 a {\em cross-preserving} 1-plane embedding  of $\gamma$
 when the same set of edge pairs makes crossings, i.e.,
  $\chi=\chi'$.
In other words, the planarization ${\cal G}(G,\gamma')$ is
another plane embedding of   ${\cal G}(G,\gamma)$ such that
the alternating order of edges incident to each crossing-vertex $c\in \chi$
is preserved.

\begin{figure}[htbp]
  \centering
 \includegraphics[width=.80\columnwidth]{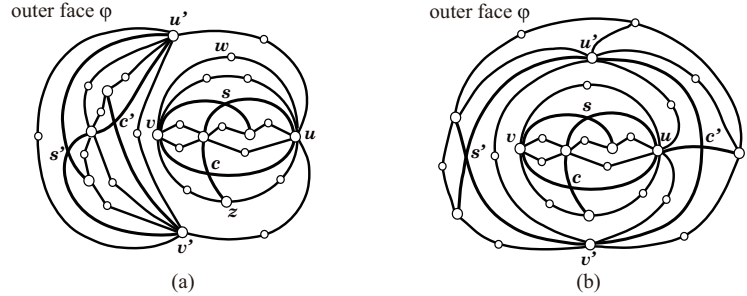}
  \caption{Circular instances $(G,\gamma)$ with a cut-vertex $u$
  of $\mathcal{G}$, where the crossing edges
     are   depicted by slightly thicker lines:
  (a) hard B-cycles $C=(u,c,v,s)$ and $C'=(u',c',v',s')$,
  (b) hard B-cycle $C=(u,c,v,s)$ and a nega-cycle $C'=(u',c',v',s')$ whose reversal is a hard B-cycle,
   where  vertices $u,v,u',v'\in V$   and crossings $c,s,c',s'\in \chi$.
}
  \label{fi:new-examplesA}
\end{figure}

To eliminate the additional constraint on the rotation system on each crossing-vertex $c\in \chi$,
we introduce ``circular instances.''
We call an instance $(G,\gamma)$ of 1-plane embedding {\em circular}
when for each crossing $c\in \chi$,
the four end-vertices of the two crossing edges $u_1u_3$ and $u_2u_4$ that create $c$
(where $u_1,u_2, u_3$ and $u_4$  appear in the clockwise order around $c$)
are contained in a  cycle $Q_c=(u_1,w_1^c,u_2,w_2^c,u_3,w_3^c,u_4,w_4^c)$ of
eight crossing-free edges for some vertices $w_i^c$, $i=1,2,3,4$ of degree 2,
as shown in Fig.~\ref{fi:BW-configuration}(c).
By definition, $c$ and each  $w_i^c$ not necessarily appear along
the same facial cycle in the planarization $\mathcal{G}(G,\gamma)$.
For example,
path $(v,w,u)$ is part of such a cycle $Q_s$ for the crossing $s$ in
the circular instance in Fig.~\ref{fi:new-examplesA}(a),
but $c$ and $w$ are not on the same facial cycle
in the planarization.

A given instance can be easily converted into a circular instance
by  augmenting the end-vertices of each pair  of crossing edges as follows.
In the plane graph,  ${\cal G}(G,\gamma)$,
for each crossing-vertex $c\in \chi$ and
its neighbors $u_1,u_2, u_3$ and $u_4$ that appear in the clockwise order around $c$,
we add a new vertex $w_i^c$, $i=1,2,3,4$ and
eight new edges $u_i w_i^c$ and $w_i^c u_{i+1}$, $i=1,2,3,4$ (where $u_5$ means $u_1$)
to form a cycle $Q_c$ of length 8 whose interior contains
no other vertex  than $c$.

Let $H$ be the resulting  graph augmented from $G$,
and let $\Gamma$ be the resulting   1-plane embedding of $H$ augmented from $\gamma$.
Note that $|V(H)|\leq  |V(G)|+4|\chi|$ holds.
 We easily see that
  if  $\gamma$  admits an SLD  cross-preserving embedding $\gamma'$
 then   $\Gamma$ admits an SLD  cross-preserving embedding $\Gamma'$.
 This is because a  straight-line drawing $D_{\gamma'}$ of $\gamma'$
 can be changed into a straight-line drawing $D_{\Gamma'}$ of
 some cross-preserving embedding $\Gamma'$ of  $\Gamma$
   by placing the newly introduced  vertices $w_i^c$
 within the region  sufficiently close to the position of $c$.
 We here see that
cycle $Q_c$ can be drawn by straight-line segments
 without intersecting with other straight-line segments in $D_{\gamma'}$.

Note that
the instance $(G,\gamma')$ remains circular
for any cross-preserving embedding $\gamma'$ of $\gamma$.
In the rest of paper,
let $(G,\gamma)$ stand for a circular  instance
 $(G=(V,E),\gamma=(\chi,\rho ,{\varphi} ) )$ with $n\geq 3$ vertices  and
  let $\mathcal{G}$ denote its planarization $\mathcal{G}(G,\gamma)$.
Fig.~\ref{fi:new-examplesA} shows examples of circular instances $(G,\gamma)$,
where the vertex-connectivity of $\mathcal{G}$ is 1.

 As an important property of a circular  instance,
 the subgraph $G_{(0)}$ with crossing-free edges is a spanning subgraph of $G$
 and  the four end-vertices of any two crossing edges are contained in
 the same block of the graph $G_{(0)}$.
 The biconnectivity is necessary to detect certain types of cycles
 by applying Lemma~\ref{le:cycle-inclusion}.

\subsection{Candidate Cycles, B/W Cycle, Posi/Nega Cycle, Hard/Soft Cycle}\label{sec:critical cycles}

For a circular instance $(G,\gamma)$, finding a cross-preserving embedding of $\gamma$
is effectively equivalent to finding another plane embedding of $\mathcal{G}$
so that all the current B- and W-configurations are eliminated
and no new B- or W-configurations are introduced.
To detect the cycles that can be the boundary of a B- or W-configuration
in changing the plane embedding of $\mathcal{G}$,
we categorize  cycles containing crossing vertices in $\mathcal{G}$.

A {\em candidate posi-cycle} (resp., {\em candidate nega-cycle})   in $\mathcal{G}$ is defined to
be   a cycle  $C=(u,c,v)$ or $C=(u,c,v,s)$   in $\mathcal{G}$
with $u,v\in V$ and $c,s\in \chi$ such that
 the interior (resp., exterior) of $C$ does not contain
a crossing-free edge  $uv\in E$ and
  any other crossing vertex $c'$  adjacent  to both $u$ and $v$.

\begin{figure}[htbp]
  \centering
 \includegraphics[width=.90\columnwidth]{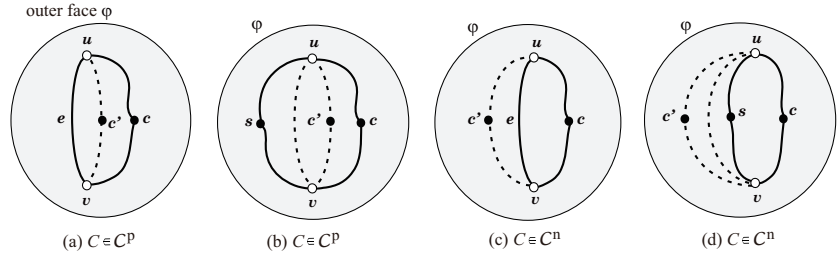}
  \caption{Candidate posi- and nega-cycles $C=(u,c,v)$ and
  $C=(u,c,v,s)$ in $\mathcal{G}$, where white circles represent
  vertices in $V$ while black ones represent crossings in $\chi$:
(a)   candidate posi-cycle of length 3,
(b)  candidate posi-cycle of length 4,
(c)  candidate nega-cycle of length 3, and
(d)  candidate nega-cycle of length 4.
}
  \label{fi:candidate-cycle_A}
\end{figure}

Fig.~\ref{fi:candidate-cycle_A}(a)-(b) and (c)-(d) illustrate
candidate posi-cycles and candidate nega-cycles, respectively.
Let $\mathcal{C}^{\mathrm{p}}$ and $\mathcal{C}^{\mathrm{n}}$ be the sets of
 candidate posi-cycles and candidate nega-cycles, respectively.
By definition we see that  the set
$\mathcal{C}^{\mathrm{p}}\cup \mathcal{C}^{\mathrm{n}}
\cup\{C_f\mid f\in F(\gamma)\}$  is inclusive,
and hence $|\mathcal{C}^{\mathrm{p}}\cup \mathcal{C}^{\mathrm{n}}
\cup\{C_f\mid f\in F(\gamma)\}|=O(n)$.

\medskip
A candidate posi-cycle $C$ with $C=(u,c,v)$   (resp.,  $C=(u,c,v,s)$) is called
a {\em B-cycle}  if \\
{\bf (a)-(B):}   the exterior of $C$ contains no   vertices in $V-\{u,v\}$ adjacent to $c$
    (resp., contains exactly one vertex in $V-\{u,v\}$ adjacent to $c$ or $s$).

\begin{figure}[htbp]
  \centering
 \includegraphics[width=0.99\columnwidth]{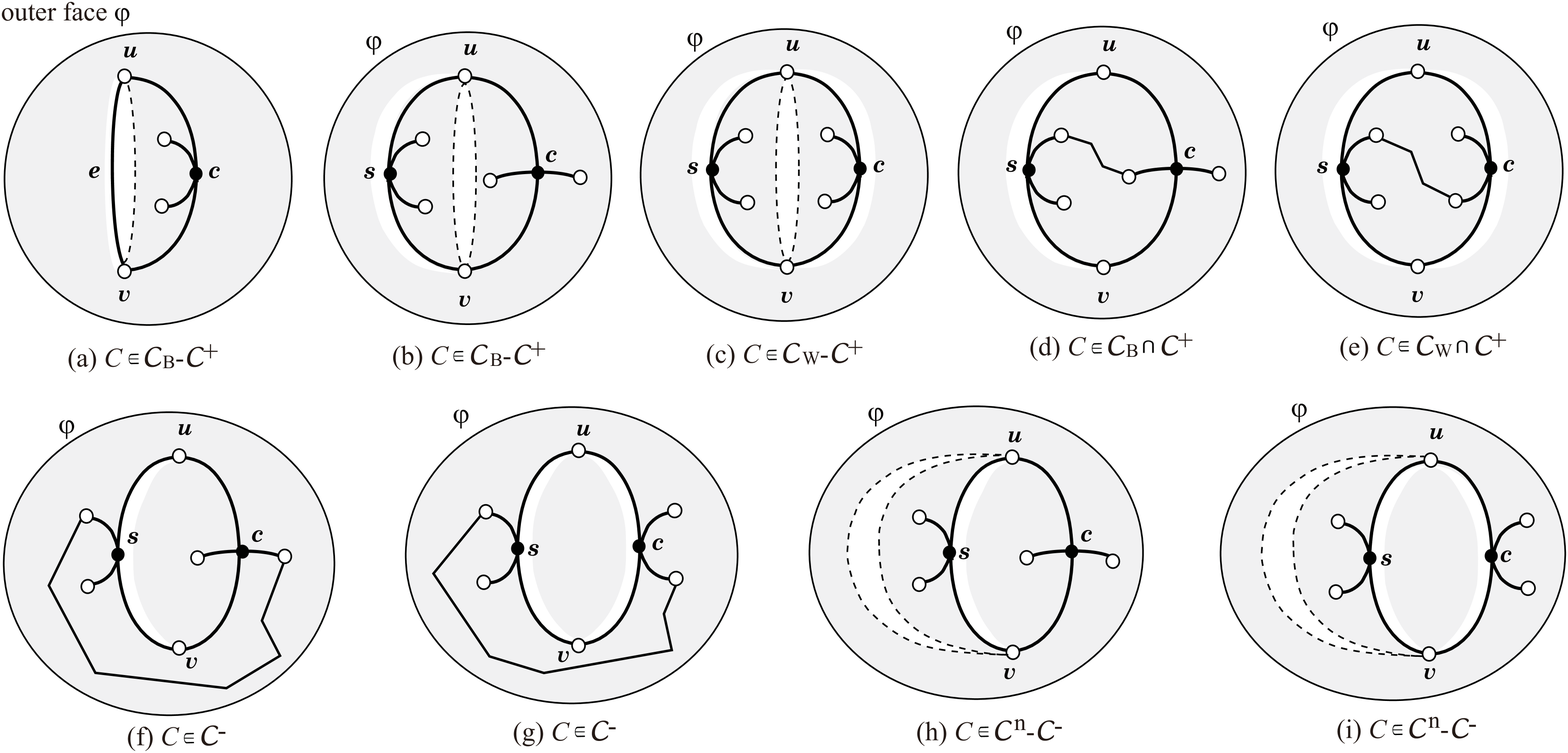}
  \caption{Illustration of types of cycles $C=(u,c,v)$ and
  $C=(u,c,v,s)$ in $\mathcal{G}$, where white circles represent
  vertices in $V$ while black ones represent crossings in $\chi$:
(a)  B-cycle of length 3, which is always soft,
(b)  soft B-cycle of length 4,
(c)  soft W-cycle,
(d)  hard B-cycle of length 4,
(e)  hard W-cycle,
(f)  nega-cycle whose reversal is a hard B-cycle,
(g)  nega-cycle whose reversal is a hard W-cycle,
(h)  candidate nega-cycle of length 4 that is not a nega-cycle whose reversal is a hard B-cycle,
and
(i)  candidate nega-cycle of length 4 that is not a nega-cycle whose reversal is a hard W-cycle.
}
  \label{fi:candidate-cycle_B}
\end{figure}

\medskip
Note that $uv\in E$ when  $C=(u,c,v)$ is a B-cycle,
as shown in Fig.~\ref{fi:candidate-cycle_B}(a).
Fig.~\ref{fi:candidate-cycle_B}(b) and (d) illustrate
the other types of B-cycles.

A candidate posi-cycle   $C=(u,c,v,s)$   is called  a {\em W-cycle}  if \\
{\bf (a)-(W):} 
 the exterior of $C$ contains no
  vertices in $V-\{u,v\}$ adjacent to  $c$ or $s$.

\medskip
Fig.~\ref{fi:candidate-cycle_B}(c) and (e) illustrate W-cycles.

Let $\mathcal{C}_{\mathrm{W}}$ (resp.,  $\mathcal{C}_{\mathrm{B}}$)
be the set of W-cycles (resp., B-cycles) in $\gamma$.
Clearly a W-cycle (resp., B-cycle) gives rise to a W-configuration
(resp., B-configuration).
Conversely, by choosing  a W-configuration
(resp., B-configuration) so that the interior is minimal,
we obtain a W-cycle (resp., B-cycle).
Hence we observe that the current embedding $\gamma$ admits a straight-line drawing
if and only if $\mathcal{C}_{\mathrm{W}}= \mathcal{C}_{\mathrm{B}}=\emptyset$.

\medskip
A W- or B-cycle $C$   is called {\em hard} if \\
{\bf (b):} length of $C$ is 4, and
  the interior  of $C=(u,c,v,s)$  contains no inner face $f$ whose facial cycle $C_f$
contains both vertices $u$ and $v$, i.e., some path connects
$c$ and $s$ without passing through $u$ or $v$.

\medskip

On the other hand,
a W- or B-cycle $C=(u,c,v,s)$ of length 4 that does not satisfy condition (b)
 or a  B-cycle of  length 3 is called {\em soft}.
We also call a hard B- or W-cycle a {\em posi-cycle}.

Fig.~\ref{fi:candidate-cycle_B}(d) and (e) illustrate
a hard B-cycle and a hard W-cycles, respectively, whereas
Fig.~\ref{fi:candidate-cycle_B}(a) and (b) (resp., (c)) illustrate
soft B-cycles (resp., a soft W-cycle).

A cycle  $C=(u,c,v,s)$ is called a {\em nega-cycle} if it becomes
a posi-cycle when an inner face in the interior of $C$ is chosen as the outer face.
In other words,  a   nega-cycle  is
a candidate nega-cycle  $C=(u,c,v,s)$ of length 4 that satisfies
the following conditions (a') and (b'),
where (a')  (resp., (b')) is obtained
 from the above  conditions (a)-(B) and (a)-(W)   (resp., (b))
by exchanging the roles of ``interior'' and  ``exterior'': \\
{\bf (a'):} 
  the interior of $C$ contains at most one
  vertex in $V-\{u,v\}$ adjacent to $c$ or $s$;  and \\
{\bf (b'):}  the exterior of $C$ contains no face $f$ whose facial cycle $C_f$
contains both vertices $u$ and $v$.

\medskip%
Fig.~\ref{fi:candidate-cycle_B}(f) and (g) illustrate nega-cycles, whereas
Fig.~\ref{fi:candidate-cycle_B}(h) and (i) illustrate
candidate nega-cycles that are not nega-cycles.

Let   $\mathcal{C}^+$  (resp., $\mathcal{C}^-$)
denote the set of   posi-cycles (resp.,  nega-cycles) in $\gamma$.
 By definition, it holds that $\mathcal{C}^+\subseteq
   \mathcal{C}_{\mathrm{W}}\cup \mathcal{C}_{\mathrm{B}}\subseteq  \mathcal{C}^{\mathrm{p}}$
 and  $\mathcal{C}^-\subseteq \mathcal{C}^{\mathrm{n}}$.

\subsection{Forbidden Cycle Pairs}\label{sec:forbidden-cycle}

We define
a forbidden configuration that characterizes
1-plane embeddings, which cannot be re-embedded into   SLD ones.
 A {\em forbidden cycle pair} is defined to be a pair $\{C,C'\}$
  of a posi-cycle $C=(u,c,v,s)$ and
  a posi- or nega-cycle $C'=(u',c',v',s')$ in $\mathcal{G}$
  with $u,v,u',v'\in V$ and $c,s,c',s'\in \chi$
  to which $\mathcal{G}$ has a $u,u'$-path $P_1$ and a $v,v'$-path $P_2$
  such that:
  \begin{enumerate}
  \item[(i)] when  $C'\in \mathcal{C}^+$, paths $P_1$ and $P_2$ are
  in the exterior of $C$ and $C'$, i.e.,
   $V(P_1)-\{u,u'\},V(P_2)-\{v,v'\}\subseteq V_{\mathrm{ex}}(C)\cap V_{\mathrm{ex}}(C')$,
where $C$ and $C'$ cannot have any common inner face; and
 \item[(ii)]  when  $C'\in \mathcal{C}^-$,
  paths $P_1$ and $P_2$ are
  in the exterior of $C$ and the interior of $C'$, i.e.,
   $V(P_1)-\{u,u'\},V(P_2)-\{v,v'\}\subseteq V_{\mathrm{ex}}(C)\cap V_{\mathrm{in}}(C')$,
where $C$ is enclosed by $C'$.
  \end{enumerate}
  
In (i) and (ii), $P_1$ and  $P_2$ are not necessary disjoint, and
possibly one of them consists of a single vertex, i.e., $u=u'$ or $v=v'$.

\begin{figure}[htbp]
  \centering
 \includegraphics[width=.80\columnwidth]{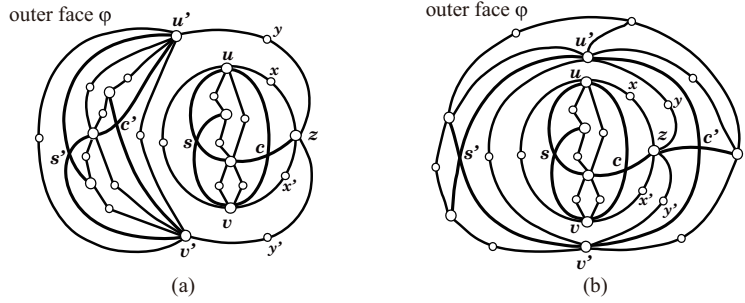}
  \caption{Illustration of circular instances $(G,\gamma)$
 with a cut-vertex $z$   of $\mathcal{G}$, where the crossing edges
     are   depicted by slightly thicker lines:
  (a) forbidden cycle pair with  hard B-cycles $C=(u,c,v,s)$ and $C'=(u',c',v',s')$
  (b) forbidden cycle pair with a  hard B-cycle $C=(u,c,v,s)$ and a nega-cycle $C'=(u',c',v',s')$
     whose reversal is a hard B-cycle,
   where  vertices $u,v,u',v'\in V$   and crossings $c,s,c',s'\in \chi$.
}
  \label{fi:new-examplesB}
\end{figure}

The pair  of cycles $C$ and $C'$ in  Fig.~\ref{fi:new-examplesB}(a)
(resp.,  Fig.~\ref{fi:new-examplesB}(b))
is a forbidden cycle pair, because
there is a pair of  a $u,u'$-path $P_1=(u,x,z,y,u')$ and a $v,v'$-path $P_2=(v,x',z,y',v')$
that satisfy the above conditions (i) (resp., (ii)).
Note that the pair of cycles $C$ and $C'$ in  Fig.~\ref{fi:new-examplesA}(a)-(b)
is not forbidden cycle pair, because there are no such paths.

Our main result of this paper is as follows.

\begin{theorem}\label{th:main}
 A circular instance  $(G,\gamma)$
 admits an SLD cross-preserving embedding
if and only if it has no forbidden cycle pair.
Finding an SLD cross-preserving embedding of $\gamma$ or
 a forbidden cycle pair in  $\mathcal{G}$  can be computed in linear time.
\end{theorem}

{\bf Proof of necessity: } The necessity of the theorem
follows from the next lemma.

For a cycle $C=(u,c,v,s)\in \mathcal{C}^+$ (resp., $\mathcal{C}^-$)
with $u,v\in V$ and $c,s\in \chi$ in
$\mathcal{G}$,
we call a vertex  $z\in V$  an {\em in-factor} of $C$
 if
  the exterior of $C\in \mathcal{C}^+$
(resp., the interior of $C\in \mathcal{C}^-$)
has   a $z,u$-path $P_{z,u}$ and
a $z,v$-path $P_{z,v}$, i.e.,
$V(P_{z,u}-\{u\})\cup V(P_{z,v}-\{v\})$
is in $V_\mathrm{ex}(C)$ (resp., $V_\mathrm{in}(C)$).
Paths $P_{z,u}$ and   $P_{z,v}$   are not necessarily disjoint.

\begin{lemma} \label{le:forbidden-cycle-pair}
Given $\mathcal{G}=\mathcal{G}(G,\gamma)$,
let   $\gamma'$ be a cross-preserving embedding of $\gamma$.
Then:  \vspace{-1mm}
\begin{enumerate}
\item[{\rm (i)}]
Let $z\in V$ be an in-factor of a cycle $C\in \mathcal{C}^+\cup\mathcal{C}^-$ in $\mathcal{G}$.
Then cycle $C$ is a posi-cycle $($resp., a nega-cycle$)$ in  $\mathcal{G}(G,\gamma')$
 if and only if
 $z$ is  in the exterior $($resp., interior$)$ of $C$ in $\gamma'$;
\item[{\rm (ii)}]
  For  a forbidden cycle pair $\{C,C'\}$,
 one of $C$ and $C'$ is  a posi-cycle in $\mathcal{G}(G,\gamma')$
 $($hence any cross-preserving embedding of $\gamma$
 contains a B- or W-configuration
 and $(G,\gamma)$ admits no SLD cross-preserving embedding$)$.
\end{enumerate}
\end{lemma}

{\bf Proof of sufficiency: }
In the rest of paper, we prove
 the sufficiency of Theorem~\ref{th:main}
 by designing a linear-time algorithm that constructs
an SLD cross-preserving embedding of an instance without
a forbidden cycle pair.

\section{Biconnected Case}\label{sec:biconnected}

In this section,   $(G,\gamma)$ stands for  a circular instance
 such that the  vertex-connectivity of the plane graph $\mathcal{G}$  is at least 2.
In a biconnected graph $\mathcal{G}$,
any two posi-cycles $C=(u,c,v,s),$ $C'=(u',c',v',s')\in \mathcal{C}^+$with $u,v,u',v'\in V$
give a forbidden cycle pair if they do not share an inner face,
because there is a pair of $u,u'$-path and $v,v'$-path in the exterior of $C$ and $C'$.
Analogously any pair of a posi-cycle $C$ and a nega-cycle $C'$
such that $C'$ encloses $C$ is also a  forbidden cycle pair
  in a biconnected graph $\mathcal{G}$.

To detect such a forbidden pair in  $\mathcal{G}$ in linear time,
we first compute the sets
$\mathcal{C}_{\mathrm{p}}$,  $\mathcal{C}_{\mathrm{n}}$,
 $\mathcal{C}_{\mathrm{W}}$,
 $\mathcal{C}_{\mathrm{B}}$, $\mathcal{C}^+$ and  $\mathcal{C}^-$ in $\gamma$
 in linear time by using the inclusion-forest from Lemma~\ref{le:cycle-inclusion}.

\invis{
\begin{figure}[htbp]
  \centering
 \includegraphics[width=.75\columnwidth]{compute-candidate-cycle.eps}
  \caption{Illustration of $u,v$-paths   of length 1 or 2 generated by
  an interval $I=[t_1,t_2,\ldots,t_q]$, where white circles represent
  vertices in $V$ while black ones crossings in $\chi$:
(a) The case where $uv$ is not a crossing-free edge,
(b) The case where $uv$ is  a crossing-free edge,  where
$I$ contains a tuple $t_k=t(e)$.
}
  \label{fi:compute-candidate-cycle}
\end{figure}
}

\begin{lemma} \label{le:candidate-cycle}
Given $(G,\gamma)$,
the following in (i)-(iv) can be computed in $O(n)$ time. \vspace{-1mm}
\begin{enumerate}
\item[{\rm (i)}]
The sets $\mathcal{C}_{\mathrm{p}}$, $\mathcal{C}_{\mathrm{n}}$ and
the inclusion-forest $\mathcal{I}$ of
$\mathcal{C}_{\mathrm{p}}\cup \mathcal{C}_{\mathrm{n}}
\cup\{C_f\mid f\in F(\gamma)\}$;
\item[{\rm (ii)}]
The sets $\mathcal{C}_{\mathrm{W}}$ and $\mathcal{C}_{\mathrm{B}}$;
\item[{\rm (iii)}]
The sets $\mathcal{C}^+$, $\mathcal{C}^-$
and the inclusion-forest $\mathcal{I}^*$ of
$\mathcal{C}^+\cup \mathcal{C}^-$; and
\item[{\rm (iv)}]
A set $\{f_C \mid
 C\in (\mathcal{C}_{\mathrm{W}}\cup\mathcal{C}_{\mathrm{B}}) -\mathcal{C}^+\}$
such that
 $f_C$ is an inner face in the interior of
a soft B- or W-cycle  $C$ with $V(C_f)\supseteq V(C)$.
\end{enumerate}
\end{lemma}

 Given $(G,\gamma)$,
a face $f\in F(\gamma)$ is called {\em admissible}
 if all posi-cycles enclose  $f$ but
no nega-cycle encloses $f$.
Let  $A(\gamma)$ denote the set of all
 admissible faces  in $ F(\gamma)$.

\begin{lemma} \label{le:admissible_face}
 Given $(G,\gamma)$,
it holds $A(\gamma)\neq \emptyset$
 if and only if no forbidden cycle pair exists in $\gamma$.
A forbidden cycle pair, if one exists, and $A(\gamma)$  can be
   obtained in $O(n)$ time.
\end{lemma}

By the lemma, if $(G,\gamma)$ has no forbidden cycle pair, i.e., $A(\gamma)\neq\emptyset$,
then any new embedding obtained from $\gamma$ by changing the outer face with
a face in $A(\gamma)$ is a cross-preserving embedding of $\gamma$ which
has no hard B- or W-cycle.

\subsection{Eliminating Soft B- and W-cycles}\label{sec:eliminate-soft}

Suppose that we are given a circular instance $(G,\gamma)$ such that
$\mathcal{G}$ is biconnected and $\mathcal{C}^+=\emptyset$.
We now show how to eliminate all soft B- and W-cycles in $\mathcal{G}$
in linear time
 using the inclusion-forest from Lemma~\ref{le:cycle-inclusion} and the spindles from Lemma~\ref{le:spindle}.

\begin{lemma} \label{le:eliminate-soft}
Given $(G,\gamma)$ with  $\mathcal{C}^+=\emptyset$,
  there exists an  SLD cross-preserving embedding
 $\gamma'=(\chi,\rho',{\varphi}' )$  of $\gamma$ such that
 $V(C_{\varphi'})\supseteq V(C_{\varphi})$ for the  facial cycle
 $C_{\varphi}$ $($resp., $C_{\varphi'})$ of the outer face  $\varphi$
 $($resp., $\varphi')$,
  which can be constructed in $O(n)$ time.
\end{lemma}

Given an instance $(G,\gamma)$ with a biconnected graph $\mathcal{G}$, we
can test whether it has either
 a forbidden cycle pair or an admissible face by Lemmas~\ref{le:candidate-cycle} and ~\ref{le:admissible_face}.
In the former, it cannot have an  SLD cross-preserving embedding by Lemma~\ref{le:forbidden-cycle-pair}.
In the latter, we can eliminate all hard B- and W-cycles
by choosing an admissible face as a new outer face,
and then  eliminate all soft B- and W-cycles
by a flipping procedure based on Lemma~\ref{le:eliminate-soft}.
All the above can be done in linear time.

\medskip
To treat the case where the vertex-connectivity of $\mathcal{G}$ is 1 in the next section,
we now characterize 1-plane embeddings that can have an  SLD cross-preserving embedding
such that a specified vertex appears along the outer boundary.
 For a vertex $z\in V$ in a graph $G$,
 we call a 1-plane embedding $\gamma$ of $G$  {\em $z$-exposed}
 if   vertex $z$   appears  along the outer boundary of $\gamma$.
 We call $(G,\gamma)$ {\em $z$-feasible}
 if it admits a  $z$-exposed SLD cross-preserving embedding $\gamma'$ of $\gamma$.

\begin{lemma} \label{le:v-feasibility}
Given $(G,\gamma)$  such that  $A(\gamma)\neq \emptyset$,
let $z$ be a vertex in $V$.
Then: \vspace{-1mm}
\begin{enumerate}
\item[{\rm (i)}]
The following conditions are equivalent: \\
{\rm (a)}
$\gamma$ admits no $z$-exposed SLD cross-preserving embedding; \\
{\rm (b)}
$A(\gamma)$ contains no  face $f$ with $z\in V(C_f)$; and \\
{\rm (c)}
$\mathcal{G}$ has a posi- or nega-cycle $C$ to which $z$ is an in-factor;
\item[{\rm (ii)}]
A $z$-exposed SLD cross-preserving embedding  or
 a posi- or nega-cycle $C$ to which $z$ is an in-factor can be
   computed in $O(n)$ time.
\end{enumerate}
\end{lemma}

\section{One-connected Case}\label{sec:connectivity-one}

In this section,   we   prove  the sufficiency of Theorem~\ref{th:main}
 by designing a linear-time algorithm claimed in the theorem.
Given a  circular instance $(G,\gamma)$,
where  $\mathcal{G}$ may be disconnected,
obviously we only need to test each connected component of  $\mathcal{G}$
separately to find a forbidden cycle pair.
Thus  we first consider  a  circular instance $(G,\gamma)$
such that the  vertex-connectivity of $\mathcal{G}$  is 1; i.e., $\mathcal{G}$ is connected
 and has some cut-vertices.

A block $B$ of $\mathcal{G}$ is a maximal biconnected subgraph of $\mathcal{G}$.
For a biconnected graph $\mathcal{G}$,
we already know how to find a forbidden cycle pair or
an SLD cross-preserving embedding from the previous section.
For a trivial block $B$ with $|V(B)|=2$,  there is nothing to do.
If some block $B$   of $\mathcal{G}$ with $|V(B)|\geq 3$
contains a forbidden cycle pair, then $(G,\gamma)$ cannot admit
any SLD cross-preserving embedding by Lemma~\ref{le:forbidden-cycle-pair}.

We now  observe that
 $\mathcal{G}$ may contain a forbidden cycle pair even if
 no single block of $\mathcal{G}$ has a forbidden cycle pair.

\begin{lemma}\label{le:block-infeasible}
For a circular instance $(G,\gamma)$
  such that the  vertex-connectivity of $\mathcal{G}$ is 1,
  let $B_1$ and $B_2$ be blocks  of $\mathcal{G}$ and
  let $P_{1,2}$ be a $z_1,z_2$-path of $\mathcal{G}$ with the minimum number of edges,
where $V(B_i)\cap V(P_{1,2})=\{z_i\}$ for each $i=1,2$.
If $\gamma|_{B_i}$ has a posi- or nega-cycle $C_i$ to which $z_i$ is an in-factor for each $i=1,2$,
then $\{C_1,C_2\}$ is a forbidden cycle pair in  $\mathcal{G}$.
\end{lemma}

For a linear-time implementation, we do not apply the lemma for all pairs of blocks in $\mathcal{B}$.
A block of $\mathcal{G}$ is called a {\em leaf block}
 if it contains only one cut-vertex of $\mathcal{G}$,
where we denote the cut-vertex in a leaf block $B$ by $v_B$.
Without directly searching for a forbidden cycle pair in $\mathcal{G}$,
we use  the next lemma to reduce a given embedding
by repeatedly removing  leaf blocks.

\begin{lemma}\label{le:good_leaf_block}
For a circular instance
 $(G,\gamma)$ such that the  vertex-connectivity of $\mathcal{G}=\mathcal{G}(G,\gamma)$    is 1 and
  a leaf block $B$ of $\mathcal{G}$ such that $\gamma|_{B}$ is $v_B$-feasible,
  let $H=G-(V(B)-\{v_B\})$ be the graph obtained
by removing the vertices in $V(B)-\{v_B\}$.
  Then \vspace{-1mm}
\begin{enumerate}
\item[{\rm (i)}]
   The instance $(H,\gamma|_{H})$ is circular; and
\item[{\rm (ii)}]
If  $(H,\gamma|_{H})$ admits an SLD cross-preserving embedding
$\gamma^*_{H}$,
 then an SLD cross-preserving embedding $\gamma^*$ of $\gamma$ can be obtained
  by placing a $v_B$-exposed SLD cross-preserving embedding  $\gamma^*_B$ of  $\gamma|_{B}$
  within a space next to the cut-vertex $v_B$ in $\gamma^*_{H}$.
\end{enumerate}
\end{lemma}

Given a circular instance $(G,\gamma)$ such that
   $\mathcal{G}=\mathcal{G}(G,\gamma)$ is connected,
   an algorithm {\bf Algorithm Re-Embed-1-Plane} for Theorem~\ref{th:main} is designed
   by the following three steps.

The first step tests   whether $\mathcal{G}$ has a block $B$   such that
 $\gamma|_B$ has a forbidden cycle pair, based on Lemma~\ref{le:v-feasibility}.
  If one exists, the algorithm  outputs     a forbidden cycle pair
  and halts.

After the first step, no block has a forbidden cycle pair.
In the current circular instance $(G,\gamma)$, one of the following holds: \\
(i) the number of blocks in $\mathcal{G}$ is at least two
and there is at most one leaf block  $B$  such that
$\gamma|_B$ is not $v_B$-feasible;   \\
(ii) $\mathcal{G}$ has  two leaf blocks $B$ and $B'$ such that
$\gamma|_B$ is not $v_B$-feasible and $\gamma|_{B'}$ is not $v_{B'}$-feasible; and \\
(iii) the number of blocks in $\mathcal{G}$ is at most one.

\medskip

In (ii),
$v_B$ is an in-factor of a cycle $C$ in $\gamma|_{B}$
and  $v_{B'}$ is an in-factor of a cycle $C'$ in $\gamma|_{B'}$
by Lemma~\ref{le:v-feasibility},
and we obtain a forbidden cycle pair $\{C,C'\}$
by Lemma~\ref{le:block-infeasible}.
Otherwise if (i) holds, then we can remove all leaf blocks  $B$  such that
$\gamma|_B$ is not $v_B$-feasible  by Lemma~\ref{le:good_leaf_block}.
The second step keeps removing  all leaf blocks $B$  such that
$\gamma|_B$ is not $v_B$-feasible
until (ii) or (iii) holds to the resulting embedding.
If (i) occurs, then the algorithm  outputs  a forbidden cycle pair
  and halts.

When all the blocks of $\mathcal{G}$ can be removed successfully,
say in an order of $B^1,B^2,\ldots,$ $ B^m$,
the third step constructs an embedding with no B- or W-cycles
by starting with such an SLD embedding of $B^m$
and by adding an SLD embedding of $B^i$ to the current embedding
in the order of $i=m-1,m-2,\ldots,1$.
By Lemma~\ref{le:good_leaf_block}, this results in an   SLD cross-preserving embedding
of the input instance $(G,\gamma)$.

\medskip
Note that we can obtain an SLD cross-preserving embedding $\gamma^*_{H^1}$ of $\gamma$
 in the third step when the first and second step did not find
 any  forbidden cycle pair.
Thus the algorithm finds either  an SLD cross-preserving embedding  of $\gamma$
or a  forbidden cycle pair.
  This proves  the sufficiency of Theorem~\ref{th:main}.

By the time complexity result from Lemma~\ref{le:v-feasibility},
we see that the algorithm can be implemented in linear time.



\begin{thebibliography}{99}


\bibitem{Franz}
 Auer, C.,  Bachmaier, C., Brandenburg, F.~J. , Glei{\ss}ner, A.,
Hanauer,  K., Neuwirth D., Reislhuber J.:
Outer 1-Planar Graphs,  Algorithmica, 74(4), 1293--1320 (2016)


\bibitem{BT}
Di Battista G., Tamassia R.:
On-line Planarity Testing,
{SIAM J. on Comput.}, 25(5),   956--997  (1996)

\bibitem{Pascal}
 Eades, P., Hong S-H., Katoh N., Liotta, G.  Schweitzer P.,  Suzuki Y.:
A linear Time Algorithm for Testing Maximal 1-planarity of Graphs with a Rotation System,
Theor. Comput. Sci., 513,   65--76  (2013)

\bibitem{FM}
Fabrici I.,  Madaras T.:
The Structure of 1-planar Graphs,
Discrete Mathematics, 307(7-8),   854--865  (2007)

\bibitem{Fary}
F\'{a}ry I. :
On straight line representations of planar Graphs,
Acta Sci. Math. Szeged, 11, 229--233  (1948)

\bibitem{Bodlaender}
 Grigoriev A.,  Bodlaender H.:
Algorithms for Graphs embeddable with few crossings per edge,
Algorithmica, 49(1),   1--11  (2007)


\bibitem{GD13}
 Hong S-H., Eades P., Katoh N., Liotta G., Schweitzer  P.,  Suzuki Y.:
A linear-time Algorithm for Testing Outer-1-planarity,
Algorithmica, 72(4),  1033--1054  (2015)

\bibitem{HELP}
Hong S-H., Eades P., Liotta  G., Poon  S.:
Fary's Theorem for 1-planar Graphs,
In:
J. Gudmundsson, J., J. Mestre J.,  Viglas T.   (eds.)
 COCOON 2012
LNCS, vol. 7434,
pp. 335--346.  Springer, Heidelberg  (2013)


\bibitem{TR} Hong S-H, Nagamochi H.:
Re-embedding a 1-Plane Graph  into a  Straight-line Drawing in Linear Time,
Technical Report TR 2016-002, Department of Applied Mathematics and Physics,
Kyoto University (2016)

\bibitem{Hopcroft3}
 Hopcroft J. E., Tarjan  R. E.:
Dividing a Graph into Triconnected Components,
SIAM J. on Comput., 2,  135--158  (1973)


\bibitem{Mohar}
 Korzhik V.~P.,  Mohar B.:
Minimal Obstructions for 1-immersions and Hardness of 1-planarity Testing,
 J. Graph Theory, 72(1),   30--71 (2013)



\bibitem{PT}
 Pach J., Toth  G.:
Graphs Drawn with Few Crossings per Edge,
 Combinatorica, 17(3),  427--439 (1997)

\bibitem{Ringel}
 Ringel  G.:
Ein Sechsfarbenproblem auf der Kugel,
Abh. Math. Semin. Univ. Hamb., 29,   107--117 (1965)



\bibitem{Thomassen}
 Thomassen C.:
Rectilinear Drawings of Graphs,
J. Graph Theory, 10(3),    335--341  (1988)



\end{thebibliography}
\end{document}